\documentstyle[preprint,eqsecnum,aps]{revtex}

\begin{document}

\preprint{\vbox{
   \hbox{NORDITA-96/18 P}
   \hbox{March 1996}
   \hbox{hep-ph/9603266}}}

\title{Color-Octet $J/\psi$ Production at Low $p_\perp$}

\author{Wai-Keung Tang\cite{DOE}}
\address{Stanford Linear Accelerator Center,
Stanford University, Stanford, CA 94309}

\author{M. V\"anttinen\cite{SA}}
\address{NORDITA, Blegdamsvej 17, DK-2100 Copenhagen}

\date{\today}

\maketitle

\begin{abstract}
We study contributions from color-octet quarkonium formation
mechanisms to $J/\psi$ hadroproduction at low $p_\perp$.
We include transitions of color-octet $c\bar{c}$ states
into ``direct'' $J/\psi$ and into $\chi_{1,2}$ which decay
radiatively into a $J/\psi$. Together with earlier work, this
calculation constitutes a complete analysis of $p_\perp$-integrated
$J/\psi$ production at leading twist. We find that the
leading-twist contribution is not sufficient to reproduce the
observed production rates and polarization of the $J/\psi$ and
$\chi_{1,2}$. Hence there must exist other important
quarkonium production mechanisms at low $p_\perp$.
\end{abstract}

\section{Introduction}

The production and decays of heavy quarkonia have been widely
studied using perturbative QCD \cite{Schuler}. Due to the
large masses of the $c$ and $b$ quarks, the production or
annihilation of heavy-quark--antiquark pairs takes place at
a much shorter timescale than the formation of the bound
state. This makes it possible to factorize the transition
amplitudes. In hadroproduction reactions, also the initial-state
hadronic structure is usually expected to be factorizable from the
$Q\bar{Q}$ production dynamics in terms of single parton
distributions; in other words, quarkonium production is usually
taken to be a leading-twist process.

In the simplest approach, one takes the hadroproduction cross
section to be a fixed fraction of the integrated $Q\bar{Q}$
cross section below the open heavy flavour threshold. This is
called the semilocal duality model or color evaporation model.
This model reproduces successfully the dependence of the
cross sections on the center-of-mass energy of the collision
and on the longitudinal momentum of the
quarkonium \cite{Gavai}. However, many observables such as the
relative production rates of different charmonium states,
transverse momentum distributions, and quarkonium polarization
are not predicted. In a more sophisticated approach, a
color-singlet $Q\bar{Q}$ pair with the appropriate $J^{PC}$
quantum numbers is produced in the hard process.
The quarkonium production amplitude is then written as the
convolution of a perturbative amplitude for the production
of a \mbox{$Q\bar{Q} \left[ ^{2S+1}L_J^{(1)} \right] $} pair
(the superscript ``(1)'' stands for a color-singlet configuration)
and a nonrelativistic wave function. This model, in which only
the leading, color-singlet Fock component of the quarkonium
wave function is taken into account, is called the color singlet
model or charmonium model.

In the color singlet model, the polarization of final-state
quarkonium is determined by the perturbative dynamics of $Q\bar{Q}$
production and by the angular momentum projection in the wave
function. Since polarization is relatively insensitive to
higher-order corrections, it provides a very good probe
of the basic $Q\bar{Q}$ production mechanisms. 

Experimentally, $J/\psi$ polarization has been measured in
fixed-target $\pi^- N$ \cite{Badier,Biino,Akerlof}
and $\bar{p}N$ \cite{Akerlof} reactions. The parameter
$\alpha$ in the angular
distribution \mbox{$1 + \alpha\cos^2\theta$}
of $J/\psi$ decay dileptons in the Gottfried-Jackson frame
has been measured to be
\mbox{$\alpha = 0.028 \pm 0.004$} in $\pi^- N$ reactions
and \mbox{$\alpha = -0.115 \pm 0.061$} in $\bar{p}N$
reactions for longitudinal momentum fraction $x_F>0$ at
125 GeV \cite{Akerlof}. This corresponds to
unpolarized production.
The theoretical interpretation of
these results is complicated by the fact that the ``direct''
and $\chi_c$-decay components of $J/\psi$ production have
not been resolved in the polarization analysis. However,
the polarization of the $\psi'(2S)$ has been
measured in $\pi^- N$ reactions \cite{Heinrich}.
The $\psi'$ is also produced unpolarized, with
\mbox{$\alpha = 0.02 \pm 0.14$} for $x_F>0.25$ at 253 GeV.
Apart from an overall normalization factor, the direct
$J/\psi$ cross section is expected to be similar to the
total $\psi'$ cross section, because no significant contribution
to $\psi'$ production from the decays of higher-mass states
has been observed.
We therefore assume that the direct $J/\psi$ component is
unpolarized; the $\chi$-decay component should then also be
unpolarized.

In Ref.\ \cite{VHBT}, we calculated the leading-twist contribution
to the $p_\perp$-integrated cross section
\mbox{$\sigma(\pi^- N \rightarrow J/\psi(\lambda) + X)$},
where $\lambda$ is the helicity of the $J/\psi$,
within the color singlet model. We included contributions
from the direct mechanism \mbox{$gg \rightarrow J/\psi + g$} and
from the radiative decays
of $\chi_J$ states produced in the reactions \mbox{$gg \rightarrow \chi_2$},
\mbox{$gg \rightarrow \chi_1 g$}, \mbox{$qg \rightarrow \chi_1 q$},
and \mbox{$q\bar{q} \rightarrow \chi_1 g$}. The polarization
analysis of the decay contribution is made possible by the
electric dipole nature of the decay.
In $\psi'$ production, only the direct mechanism
needs to be taken into account.

The leading-twist, color-singlet contributions to $J/\psi$ and
$\psi'$ production turn out to be dominantly transversely polarized,
i.e.\ \mbox{$\sigma(\lambda=\pm 1) > \sigma(\lambda=0)$},
with \mbox{$\alpha \sim 0.5$} for the total $J/\psi$ and
\mbox{$\alpha \sim 0.3$} for the $\psi'$ and direct $J/\psi$
produced by pions. This is in contrast to the experimental
observation of unpolarized production. The model also fails to
reproduce the observed relative production rates \cite{AntoniazziPRL70}
of the $J/\psi$, $\chi_1$ and $\chi_2$ states. In Ref.\
\cite{VHBT}, we interpreted these discrepancies as evidence for
important higher-twist mechanisms of charmonium production.

However, one could argue that the discrepancies are due to
neglecting the higher Fock components in the quarkonium
wavefunction. These are systematically included in nonrelativistic
QCD (NRQCD), an effective field theory which has been formulated
during the recent years \cite{Lepage,BBL}. NRQCD provides an
expansion of quarkonium cross sections and decay widths in
terms of the relative velocity $v$ of the heavy quark and antiquark
in the bound state. The nonperturbative physics is factorized into
an infinite number of matrix elements that scale in a well-defined
way in $m_Q$ and $v$.

Color-octet mechanisms have recently been suggested
\cite{BraatenFleming,ChoLeibovich,ChoII} as an explanation of the
large quarkonium production rates observed at the Tevatron
$p\bar{p}$ collider \cite{CDF,D0}.
Some of these rates exceed color-singlet-model predictions
by more than an order of magnitude.

There exist color-octet mechanisms of low-$p_\perp$
quarkonium production which are of higher order in $v^2$ but
lower order in $\alpha_s$ compared to the leading color-singlet mechanisms.
They might help to reproduce the fixed-target data;
the purpose of this paper is to find out whether they do.
If not, then the fixed-target
discrepancies must be due to a breakdown of the leading-twist
approximation of $Q\bar{Q}$ production.

In a recent work \cite{TV}, we already analyzed the color-octet
contribution to $\psi'$ hadroproduction at low $p_\perp$. We found that
the leading color-octet contribution is dominantly transversely polarized.
Hence, even with the leading color-octet components
included, the unpolarized $\psi'$ production data \cite{Heinrich}
cannot be reproduced. This suggests that there are important
higher-twist $Q\bar{Q}$ production mechanisms. The $\psi'$ analysis
also applies to the direct component of $J/\psi$ production.
In the present paper, we complete our analysis by calculating the
contribution to $J/\psi$ production from the radiative decays of
$\chi_{1}$ and $\chi_{2}$ produced through color-octet
intermediate states. It will turn out that this component, too, is
dominantly transversely polarized. Furthermore,
determinations of color-octet matrix elements from the analysis
of other reactions imply that the magnitude of this component
is small compared to the data. We shall conclude that
in $J/\psi$ as well as in $\psi'$ production,
new mechanisms are likely to be important at low $p_\perp$.

\section{Summary of color-singlet $J/\psi$ production cross sections}

At leading order in $\alpha_s$ and at leading twist, the
color-singlet $Q\bar{Q}$ production subprocesses are
\begin{eqnarray}
  gg       & \rightarrow & ^1S_0, \; \mbox{}^3P_{0,2}, \label{Baier1} \\
  gg       & \rightarrow & ^3S_1+g, \; \mbox{}^3P_J+g, \label{Baier2} \\
  gq       & \rightarrow & ^3P_J+q, \label{Baier3}\\
  q\bar{q} & \rightarrow & ^3P_J+g , \label{Baier4}
\end{eqnarray}
i.e.\ they are obtained from the leading-order annihilation
subprocesses such as \mbox{$J/\psi \rightarrow ggg$} by crossing.
Note that the process \mbox{$gg \leftrightarrow \;\mbox{}^3P_1$,}
with the gluons on mass shell, is forbidden by Yang's theorem
\cite{Yang}. Within the color singlet model, the $Q\bar{Q}$
production mechanism could be different from
(\ref{Baier1}--\ref{Baier4}); it could e.g.\ be a higher-twist
process, where many partons from the same initial-state hadron
participate in the hard scattering.

The radiative decays \mbox{$\chi_{1,2} \rightarrow J/\psi + \gamma$}
are known
experimentally to be a major source of $J/\psi$ production.
They contribute 30-40\% of the total cross section
in both pion- and proton-induced reactions \cite{AntoniazziPRL70}.
The production of the $\psi'$, on the other hand, is expected to
be dominantly due to the direct subprocess (\ref{Baier2}).

The polarization of
a $^3S_1$ state like the $J/\psi$ is reflected in the
polar-angle distribution of its decay dileptons in their rest
frame. The parameter $\alpha$ in the angular
distribution \mbox{$1 + \alpha\cos^2\theta$} is related to the
polarized $J/\psi$ production cross section
\begin{equation}
  \sigma(\lambda) = \sigma_{\rm tot}\; \frac{1 + a\delta_{\lambda 0}}{3+a},
\end{equation}
where $\lambda$ is the helicity of the $J/\psi$,
in the following way:
\begin{equation}
  \alpha = \frac{d\sigma(\lambda=1) - 2 d\sigma(\lambda=0)
           + d\sigma(\lambda=-1)}{ d\sigma(\lambda=1)
           + 2 d\sigma(\lambda=0) + d\sigma(\lambda=-1)}
         = -\frac{a}{2+a} .
\end{equation}
Because of rotational and parity invariance,
\mbox{$d\sigma(\lambda=1) = d\sigma(\lambda=-1)$}.
We work in the Gottfried-Jackson frame, which is
defined as the particular quarkonium rest frame where the
beam momentum lies along the $z$ axis and the target momentum
lies in the $xz$ plane, with $p_x({\rm target}) \le 0$. At
vanishing transverse momentum of the quarkonium, the
Gottfried-Jackson frame is obtained from the laboratory
frame by a simple boost, and the choice of the $x$ axis
is immaterial.

We analyzed the polarized cross sections of $J/\psi$
production through the direct process and $\chi_J$ decays in
Ref.\ \cite{VHBT}, where we found that the color singlet
model is insufficient to explain the existing fixed-target data.
In the color singlet model, the $\chi_1$ and direct $J/\psi$
cross sections are predicted to be significantly lower
than measured relative to the $\chi_2$ cross section, which
is within a factor $K =$ 2--3 of the experimental value.
In contrast with the observed unpolarized
production, the predicted total $J/\psi$ cross section
is dominantly transversely polarized even if the various
contributions are renormalized according to the data.
The quantitative results of the calculation
of Ref.\ \cite{VHBT} are listed in Table \ref{table:all}
together with experimental data and color-octet predictions.

\section{Color-octet $J/\psi$ production cross sections}

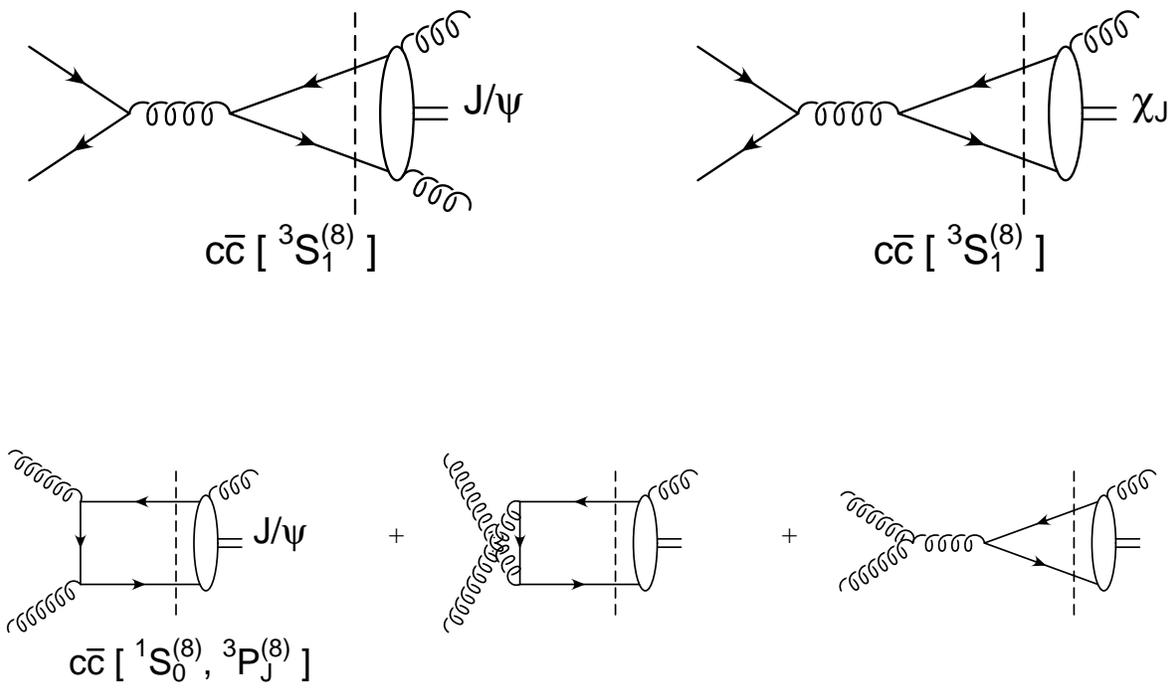
\begin{figure}
\caption{The Feynman diagrams which describe the leading color-octet
mechanisms of $J/\psi$ and $\chi_J$ production. The blob
represents a nonperturbative transition. The dashed line indicates
the color-octet intermediate state.}
\label{diagrams}
\end{figure}

As one calculates quarkonium production cross sections within the
NRQCD factorization scheme, one is using two expansions:
the perturbative expansion
of the short-distance $Q\bar{Q}$ production amplitude and the
velocity expansion of the long-distance quarkonium formation
amplitude. General rules for finding
out the power dependence of the NRQCD matrix elements on
$m_Q$ and $v$ can be found in Refs.\ \cite{Lepage,BBL}.
%
%
At leading order in perturbation theory, i.e.\ $O(\alpha_s^2)$,
and up to next-to-leading order in the velocity
expansion, the subprocesses for leading-twist $J/\psi$ production
through color-octet intermediate states are
\begin{eqnarray}
  q\bar{q} & \rightarrow & c\bar{c} \left[ \mbox{}^3 S_1 ^{(8)} \right]
                           \rightarrow J/\psi + gg,
                           \label{octet1} \\
  gg       & \rightarrow & c\bar{c} \left[ \mbox{}^1 S_0 ^{(8)} \right]
                           \rightarrow J/\psi + g,
                           \label{octet2} \\
  gg       & \rightarrow & c\bar{c} \left[ \mbox{}^3 P_J ^{(8)} \right]
                           \rightarrow J/\psi + g,
                           \label{octet3} \\
  q\bar{q} & \rightarrow & c\bar{c} \left[ \mbox{}^3 S_1 ^{(8)} \right]
                           \rightarrow \chi_J + g
                           \rightarrow J/\psi + \gamma + g.
                           \label{octet4}
\end{eqnarray}
These are illustrated by the Feynman diagrams of Fig.\ \ref{diagrams},
where the blob represents a nonperturbative transition.
Their cross sections
are proportional to the NRQCD matrix elements
\begin{eqnarray}
  \langle 0| {\cal O}_8^{J/\psi}(^3S_1) |0\rangle
    & \sim & m_c^3 v^7, \\
  \langle 0| {\cal O}_8^{J/\psi}(^1S_0) |0\rangle
    & \sim & m_c^3 v^7, \\
  \langle 0| {\cal O}_8^{J/\psi}(^3P_J) |0\rangle
    & \sim & m_c^5 v^7, \\
  \langle 0| {\cal O}_8^{\chi_J}(^3S_1) |0\rangle
    & \sim & m_c^3 v^5,
\end{eqnarray}
respectively.
Hence the direct $J/\psi$ production cross sections are proportional
to $\alpha_s^2 v^7$, and the $\chi_J$ cross section is proportional
to $\alpha_s^2 v^5$. They are to be compared with the leading color-singlet
cross sections
\begin{eqnarray}
  \sigma(gg \rightarrow \chi_2) & \sim & \alpha_s^2 v^5, \nonumber \\
  \sigma(gg \rightarrow J/\psi + g ) & \sim & \alpha_s^3 v^3, \nonumber \\
  \sigma(ij \rightarrow \chi_1+k) & \sim & \alpha_s^3 v^5.
\end{eqnarray}

The amplitude
\begin{equation}
  A \left( q\bar{q} \rightarrow c\bar{c} \left[ ^3P_J^{(8)} \right]
  \rightarrow J/\psi + g \right)
\end{equation}
is of higher order in $v^2$ than  the amplitudes of the processes
\mbox{(\ref{octet1}--\ref{octet4}),}
because the lowest-order non-perturbative transition
(single chromoelectric dipole) is forbidden by charge conjugation.
Furthermore, the amplitude
\begin{equation}
  A \left( gg \rightarrow c\bar{c} \left[ ^3S_1^{(8)} \right]
  \rightarrow \chi_J + g \right)
\end{equation}
is of higher order in $\alpha_s$ because the amplitude
\mbox{$ A \left( gg \rightarrow c\bar{c} \left[ ^3S_1^{(8)} \right] \right)$}
vanishes in the leading order. Charge conjugation or Yang's theorem
would not require this amplitude to vanish because the $c\bar{c}$
pair is not in a color-singlet state.

The contribution to \mbox{$\sigma(hN \rightarrow J/\psi(\lambda) + X)$}
from the direct production
subprocesses (\ref{octet1}--\ref{octet3}) follows immediately from
the $\psi'$ production analysis of Ref.\ \cite{TV}:
\begin{eqnarray}
  \sigma_{\rm octet} (h N \rightarrow
  {\rm direct} \; J/\psi(\lambda) + X)
  & = & O_1
        \langle 0|{\cal O}^{J/\psi}_8 (\,^3P_1 )|0 \rangle
        (3 - 2\delta_{\lambda 0})
        \nonumber \\
  &   & \mbox{}
         + O_2
        \langle 0|{\cal O}^{J/\psi}_8 (\,^1S_0 )|0 \rangle
        (1 - \delta_{\lambda 0})
        \nonumber \\
  &   & \mbox{}
        + O_3
        \langle 0|{\cal O}^{J/\psi}_8 (\,^3S_1 )|0 \rangle
        (1 - \delta_{\lambda 0})
        .
  \label{octetdirect}
\end{eqnarray}
The coefficients are
\begin{eqnarray}
  O_1 & = & \frac{5 \pi^3 \alpha_s^2}{9 M^7}
            \; \Phi_{gg/h N}(M^2/s, \mu_F), \\
  O_2 & = & \frac{5 \pi^3 \alpha_s^2}{24 M^5}
            \; \Phi_{gg/h N}(M^2/s, \mu_F), \\
  O_3 & = & \frac{8 \pi^3 \alpha_s^2}{27 M^5}
            \; \sum_q \left[
            \Phi_{q\bar{q}/h N}(M^2/s,\mu_F)
            + \Phi_{\bar{q}q/h N}(M^2/s,\mu_F)
            \right] ,
\end{eqnarray}
where
\begin{equation}
  \Phi_{ij/h N}(\tau,\mu_F)
  \equiv \int dx_1 dx_2 \; f_{i/h}(x_1,\mu_F) \; f_{j/N}(x_2,\mu_F)
  \; \delta \left( 1 - \frac{\tau}{x_1 x_2} \right)
\end{equation}
is a parton flux factor evaluated at the leading-twist factorization
scale $\mu_F$ (since the final-state gluons are taken to be soft,
the kinematics is essentially that of a $2 \rightarrow 1$ subprocess,
with \mbox{$\hat s = M^2 = 4m_c^2$}).
For ease of reference, we plot in Fig.\ \ref{flux}
the $gg$ and $q\bar{q}$ flux factors in $\pi^-$-proton, proton-proton
and antiproton-proton collisions using the GRV-LO
parton distributions \cite{GRV}. In accordance with
the existing experiments, the integral is over the region
\mbox{$x_F = x_1-x_2 > 0$.}

The polarization of the directly produced $J/\psi$ has not been
measured separately from the polarization of those from $\chi$
decays. However, the fact that the $\psi'$ are produced
unpolarized \cite{Heinrich} lets us expect that the direct $J/\psi$
component is also unpolarized. Then, as in the $\psi'$ case, only about
half of the observed direct production can be due to the
strongly transversely
polarized contribution (\ref{octetdirect}). We discussed
the derivation of a quantitative bound on a linear combination
of NRQCD matrix elements in Ref.\ \cite{TV}.

To evaluate the contribution from the process (\ref{octet4}),
we make use of the fact that both the
$c\bar{c} \left[ ^3S_1^{(8)} \right] \rightarrow \chi_J + g$
transition and the radiative decay of the $\chi_J$ are electric
dipole transitions. The necessary formulas are given in the appendix.
From the \mbox{$q\bar{q} \rightarrow c\bar{c}$} scattering amplitude,
\begin{equation}
  {\cal O}^{AB}_{q\bar{q} \rightarrow c\bar{c}}
   = \frac{4\pi\alpha_s}{(p_{\bar{q}} + p_q)^2}
     \; \bar{v}_C(p_{\bar{q}}) \; \gamma_\mu T^a_{CD} \; u_D(p_q)
     \; T_a^{AB} \gamma^\mu,
\end{equation}
we derive
\begin{eqnarray}
  \overline{\sum_a}
  \left| A \left( q\bar{q} \rightarrow
  c\bar{c} \left[ \mbox{}^3 S_1^{(8)}; S_z, a \right] \right)
  \right| ^2
  = \frac{32\pi^2 \alpha_s^2}{9}
    \, \left( 1 - \delta_{S_z 0} \right) .
  \label{octetxs}
\end{eqnarray}
The subprocess cross sections then become
\begin{eqnarray}
  \lefteqn{\sigma \left( q\bar{q} \rightarrow
           c\bar{c} \left[ \mbox{}^3 S_1^{(8)} \right] \rightarrow
           \chi_1 + g \rightarrow J/\psi(\lambda) + \gamma + g \right)}
           \nonumber \\
  & = & \frac{16\pi^3 \alpha_s^2}{27 M^5}
        \; \delta \left( 1 - \frac{M^2}{\hat s} \right)
        {\rm Br}\, (\chi_1 \rightarrow J/\psi + \gamma)
        \left\langle 0 \left| {\cal O}^{\chi_1}_8 (^3S_1) \right| 0
        \right\rangle \frac{3 - \delta_{\lambda 0}}{8}, \\
  \vspace{2ex}
  \lefteqn{\sigma \left( q\bar{q} \rightarrow
           c\bar{c} \left[ \mbox{}^3 S_1^{(8)} \right] \rightarrow
           \chi_2 + g \rightarrow J/\psi(\lambda) + \gamma + g \right)}
           \nonumber \\
  & = & \frac{16\pi^3 \alpha_s^2}{27 M^5}
        \; \delta \left( 1 - \frac{M^2}{\hat s} \right)
        {\rm Br}\, (\chi_2 \rightarrow J/\psi + \gamma)
        \left\langle 0 \left| {\cal O}^{\chi_2}_8 (^3S_1) \right| 0
        \right\rangle \frac{47 - 21\delta_{\lambda 0}}{120} .
\end{eqnarray}
Using the NRQCD result
$
   \langle 0| {\cal O}_8^{\chi_2}(^3 S_1)|0\rangle
   = (5/3) \, \langle 0| {\cal O}_8^{\chi_1}(^3 S_1)|0\rangle ,
$
valid up to corrections of relative order $v^2$,
we arrive at
\begin{eqnarray}
  \lefteqn{\sigma_{\rm octet} (h N \rightarrow \chi_J + g
  \rightarrow J/\psi(\lambda) + \gamma + g)} \nonumber \\
  & = & \sum_q \left[
        \Phi_{q\bar{q}/h N} \left( \frac{M^2}{s}, \mu_F \right)
        + \Phi_{\bar{q}q/h N} \left( \frac{M^2}{s}, \mu_F \right)
        \right] \; \frac{16\pi^3 \alpha_s^2}{27 M^5}
        \; \langle 0| {\cal O}_8^{\chi_1}(^3 S_1)|0\rangle
        \nonumber \\
  &   & \mbox{} \times \left[
        {\rm Br}\,(\chi_1 \rightarrow J/\psi + \gamma)
        \, \frac{3 - \delta_{\lambda 0}}{8}
        \; + \; \frac{5}{3} \,
        {\rm Br}\,(\chi_2 \rightarrow J/\psi + \gamma)
        \, \frac{47 - 21\delta_{\lambda 0}}{120}
        \right] .
  \label{octetchi}
\end{eqnarray}
Our expressions for the polarized cross sections,
eqs.\ (\ref{octetdirect}, \ref{octetchi}), agree with the unpolarized cross
sections given in Refs.\ \cite{ChoII,FM}.

Each of the two components in eq.\ (\ref{octetchi}) alone would correspond to
$\alpha=0.2$ and $\alpha=0.29$ in the decay angular distribution
of the $J/\psi$,
respectively. Hence this contribution,
similarly to the color-singlet contribution \cite{VHBT}
and the direct color-octet contribution of eq.\ (\ref{octetdirect}),
is dominantly
transversely polarized. Because $J/\psi$ production in $\pi N$ reactions
is observed to be unpolarized, one is again led to conclude that
there are important
$c\bar{c}$ production mechanisms beyond leading twist.

The value of the \mbox{$\langle 0| {\cal O}_8^{\chi_1}(^3 S_1)|0\rangle$}
matrix element has been obtained from the analysis of CDF data
at large $p_\perp$ \cite{ChoLeibovich,ChoII} and of $B$ decays into
$P$-wave charmonium \cite{BraatenFleming,Bdecay}. These values
actually imply that the contribution (\ref{octetchi}) is rather
small compared to the observed $J/\psi$ cross section. Using
$
  \langle 0| {\cal O}_8^{\chi_1}(^3 S_1)|0\rangle
  = (9.8 \pm 1.3) \cdot 10^{-3} \; ({\rm GeV})^3
$
\cite{ChoII}, \mbox{$\alpha_s = 0.26$} and
\mbox{$M=3.5$ GeV}, we obtain
\begin{eqnarray}
  \lefteqn{\sum_\lambda \sigma_{\rm octet}
  (h N \rightarrow \chi_{1,2} + g \rightarrow J/\psi(\lambda) + \gamma + g)}
  \nonumber \\
  & = & 4.5 \; {\rm nb}
        \sum_q \left[
        \Phi_{q\bar{q}/h N} \left( \frac{M^2}{s}, \mu_F \right)
        + \Phi_{\bar{q}q/h N} \left( \frac{M^2}{s}, \mu_F \right)
        \right]
  =
        \left\{ \begin{array}{ll}
        1.3 \; {\rm nb} & (\pi^- \; {\rm beam}) \\
        0.6 \; {\rm nb} & (p   \; {\rm beam})
        \end{array} \right.
\end{eqnarray}
at \mbox{$E_{\rm lab}(\pi) = 300$ GeV}. These numbers are more
than an order of magnitude smaller than the experimental cross
sections of 72 nb, 67 nb and 45 nb (with errors of about 25\%)
for $\pi^+$, $\pi^-$ and $p$ beams, respectively
\cite{AntoniazziPRL70}.
Hence the color-octet mechanisms cannot reproduce
the fixed-target data. On the other hand, our analysis does not
set any constraints that would contradict the color-octet description
of other reactions.

\section{Summary}

In this paper, we have considered the production of $J/\psi$
charmonium in fixed-target reactions within the
factorization scheme of nonrelativistic QCD (NRQCD) \cite{BBL}.
We have calculated the
contribution from the radiative decays, \linebreak[2]
\mbox{$\chi_{1,2} \rightarrow J/\psi + \gamma$,} of $\chi_J$
charmonia produced through intermediate color-octet $c\bar{c}$
states at leading twist. The ``direct'' color-octet component
of $J/\psi$ production
is given by our analysis of color-octet $\psi'$
production \cite{TV}. Together with our earlier evaluation
of the color-singlet component \cite{VHBT}, the present calculation
constitutes a complete analysis of leading-twist $J/\psi$
production at low $p_\perp$.
We have included contributions from nonperturbative transitions
between intermediate $c\bar{c}$ states and charmonium up to
relative order $v^4$, where $v$ is the relative velocity of the
charm quark and antiquark in the bound state. The $c\bar{c}$
production amplitudes have in each case been evaluated at the
lowest order in $\alpha_s$ allowed by the quantum numbers of
the intermediate $c\bar{c}$ state.

The results on $J/\psi$ production obtained in this paper and
Refs.\ \cite{VHBT,TV} have been collected in Table \ref{table:all}.
In the evaluation of color-octet contributions, we have used
the NRQCD matrix elements determined in Refs.\ \cite{ChoII,AFM}.

Our motivation has been to test both the NRQCD picture of the
long-distance quarkonium formation process and the leading-twist
approximation of short-distance $Q\bar{Q}$ production.
Predictions of the final-state charmonium polarization provide
a very good test of the models involved. They are relatively
insensitive to higher-order corrections in perturbation theory.
Polarization analysis of the long-distance process is made
possible by the fact that the emission or absorption of
soft gluons does not change the heavy quark spins.

We have shown that the
leading-twist contribution to $J/\psi$ production is dominantly
transversely polarized, i.e.\ \mbox{$\alpha > 0$} in the angular
distribution, \mbox{$1 + \alpha\cos^2\theta$}, of the decay
\mbox{$J/\psi \rightarrow \ell^+ \ell^-$} in the Gottfried-Jackson
frame. Actually, all but one of the individual components of the
theoretical cross section are dominantly transversely polarized.
The only exception is the small contribution from the
radiative decay of the $\chi_1$ produced
by color-singlet mechanisms, which gives \mbox{$\alpha \approx -0.15$.}
Experimentally, on the other hand, it has been observed that
$S$-wave charmonia are produced unpolarized.

Furthermore, the values of NRQCD matrix elements determined
from other reactions imply that the normalization of color-octet
contributions is small compared to the observed charmonium
cross sections.

Assuming that the NRQCD factorization scheme provides a complete
description of quarkonium formation, the reason for the
discrepancies must lie in the choice of $c\bar{c}$ production
mechanisms. Hence there should exist important
higher-twist mechanisms of $c\bar{c}$ production at small $p_\perp$.

Since the mass of the $b$ quark is significantly larger than the
$c$ quark mass, all the approximations involved in the calculation
-- perturbation theory, the velocity expansion, and the
leading-twist approximation -- are expected to work better for
bottomonium. Unfortunately, the existing bottomonium production
data is insufficient to test our predictions.

\appendix

\section{Dipole transitions}

\subsection{Electric dipole transitions}

The polarized cross section of
$J/\psi$ production via the radiative decay of a $\chi_J$
charmonium state is
\begin{eqnarray}
  \lefteqn{\sigma \left( ij \rightarrow \chi_J + X \rightarrow
           J/\psi (\lambda) + \gamma + X \right) } \nonumber \\
  & = & \frac{1}{2\hat s} \int {\rm dLips}(J/\psi,\gamma,X)
        \overline{\sum}
        \left| A \left( ij \rightarrow \chi_J + X \rightarrow
          J/\psi (\lambda) + \gamma + X \right) \right| ^2 \nonumber \\
  & = & \frac{1}{2\hat s} \int \frac{dp^2(\chi_J)}{2\pi}
        \; {\rm dLips}(\chi_J,X) \; {\rm dLips}(J/\psi,\gamma)
        \; \overline{\sum} \left| A \right| ^2,
\end{eqnarray}
where \mbox{$\hat s = (p_i+p_j)^2$} is the invariant mass of the
initial state and $\lambda$ is the helicity of the $J/\psi$.
In the limit where the momenta of the photon and final-state
light hadrons are neglected, it equals the $z$ component of the
spin of the $J/\psi$.
The transition amplitude is
\begin{eqnarray}
  \lefteqn{A \left( ij \rightarrow \chi_J + X \rightarrow
           J/\psi (\lambda) + \gamma(\mu) + X \right) } \nonumber \\
  & = & \sum_{J_z} A \left( ij \rightarrow \chi_J(J_z) + X \right)
        \; \varphi(\chi_J) \;
        A \left( \chi_J(J_z)
          \rightarrow J/\psi(\lambda) + \gamma(\mu) \right),
\end{eqnarray}
where $\varphi(\chi_J)$ is the propagator of the $\chi_J$
and $\mu$ is the helicity of the photon.
In the limit of small $\chi_J$ decay width,
\begin{equation}
  \lim_{\Gamma_{\rm tot}(\chi_J) \rightarrow 0}
  \varphi(\chi_J) \varphi^*(\chi_J)
  = \frac{\pi}{M_{\chi_J} \Gamma_{\rm tot}(\chi_J)}
    \delta(M^2_{\chi_J} - p^2_{\chi_J}).
\end{equation}
In the electric dipole approximation, the $\chi_J$ decay amplitude
is written as
\begin{eqnarray}
  A \left( \chi_J(J_z) \rightarrow
           J/\psi (\lambda) + \gamma(\mu) \right)
  & = & \sum_{L_z S_z} \langle JJ_z|L_z S_z\rangle
        \, A \left( c\bar{c}(L_z,S_z)
        \rightarrow  J/\psi (\lambda) + \gamma(\mu) \right)
        \nonumber \\
  & = & \sum_{L_z S_z} \langle JJ_z|L_z S_z\rangle
        N \delta_{S_z \lambda} \epsilon^*_{L_z}(\mu) \nonumber \\
  & = & N \langle JJ_z|J_z-\lambda,\lambda\rangle
        \epsilon^*_{J_z-\lambda}(\mu),
\end{eqnarray}
where the symbol $\delta_{S_z \lambda}$ expresses the heavy
quark spin conservation and the normalization factor is
\begin{equation}
  N = \left( \frac{24\pi M_{J/\psi} M^2_{\chi_J} \Gamma(\chi_J \rightarrow
      J/\psi + \gamma)}{M^2_{\chi_J}-M^2_{J/\psi}} \right) ^{1/2}.
\end{equation}
Using the results
\begin{eqnarray}
  {\rm dLips}(J/\psi,\gamma)
  & = & \frac{M^2_{\chi_J}-M^2_{J/\psi}}{32\pi^2  M_{J/\psi} M^2_{\chi_J}}
        \; d\Omega, \\
  \sum_\mu \int d\Omega \; \epsilon_i(\mu) \epsilon^*_j(\mu)
  & = & \frac{8\pi}{3} \delta_{ij},
\end{eqnarray}
where $d\Omega$ is a solid angle element in the charmonium rest frame,
we obtain
\begin{eqnarray}
  \lefteqn{\sigma(ij \rightarrow \chi_J + X \rightarrow
           J/\psi (\lambda) + \gamma +X)} \nonumber \\
  & = & {\rm Br}\,(\chi_J \rightarrow J/\psi + \gamma)
        \sum_{J_z} | \langle JJ_z|J_z-\lambda,\lambda\rangle  |^2 \,
        \frac{1}{2\hat s} \int {\rm dLips}(\chi_J,X)
        \overline{\sum} \left| A(ij \rightarrow \chi_J(J_z) + X) \right| ^2
        \nonumber \\
  & = & {\rm Br}\,(\chi_J \rightarrow J/\psi + \gamma)
        \sum_{J_z} | \langle JJ_z|J_z-\lambda,\lambda\rangle  |^2 \;
        \sigma(ij \rightarrow \chi_J(J_z) + X) .
  \label{electric}
\end{eqnarray}
This result was also used in the color-singlet-model calculation
of Ref. \cite{VHBT}.

\subsection{Chromoelectric dipole transitions}

Analogously with eq.\ (\ref{electric}), we write the cross section
of $\chi_J$ production via the chromoelectric dipole "decay"
of a color-octet $c\bar{c} \left[ \mbox{}^3 S_1 \right]$
state as
\begin{eqnarray}
        \sigma \left( ij \rightarrow
        c\bar{c} \left[ \mbox{}^3 S_1^{(8)} \right]
        \rightarrow \chi_J(J_z) + g \right)
  & = & \frac{1}{8(2J+1)m_c} \langle 0| {\cal O}_8^{\chi_J}(^3 S_1)|0\rangle
        \sum_{S_z} | \langle JJ_z | J_z-S_z, S_z \rangle  |^2
        \nonumber \\
  &   & \times
        \frac{\pi}{M^4} \; \delta \left( 1 - \frac{M^2}{\hat s} \right)
        \overline{\sum_a} \;
        \left| A \left( ij \rightarrow
        c\bar{c} \left[ \mbox{}^3 S_1^{(8)}; S_z, a \right] \right)
        \right| ^2 ,
        \nonumber \\
  \label{chromoelectric}
\end{eqnarray}
where \mbox{$\langle 0| {\cal O}_8^{\chi_J}(^3 S_1)|0 \rangle$}
is a matrix element of nonrelativistic QCD and
\begin{equation}
  A \left( ij \rightarrow c\bar{c}
  \left[ \mbox{}^3 S_1^{(8)}, S_z, a \right] \right)
  = \sqrt{2} \; T^a_{AB} \; {\rm Tr}
    \left[ {\cal O}^{AB}_{ij \rightarrow c\bar{c}}
    \frac{\epsilon\!\! /^*(S_z) (P\!\!\! / + M)}{2\sqrt{2}}
    \right] .
\end{equation}
${\cal O}^{AB}_{ij \rightarrow c\bar{c}}$ is the amplitude for the
perturbative process $ij \rightarrow c\bar{c}$, as given by Feynman
rules (\mbox{$A,B$} are the color indices of the heavy quark and
antiquark; the heavy-quark and antiquark spinors are truncated),
and $M=2m_c$ is the mass of the $c\bar{c}$ state.
Combining eqs. (\ref{electric}) and (\ref{chromoelectric}), we have
the final result
\begin{eqnarray}
  \lefteqn{\sigma \left( ij \rightarrow
           c\bar{c} \left[ \mbox{}^3 S_1^{(8)} \right] \rightarrow
           \chi_J + g \rightarrow J/\psi(\lambda) + \gamma + g \right) }
           \nonumber \\
  & = & {\rm Br}\,(\chi_J \rightarrow J/\psi + \gamma)
        \frac{1}{8(2J+1)m_c} \langle 0| {\cal O}_8^{\chi_J}(^3 S_1)|0 \rangle
        \sum_{J_z} | \langle JJ_z|J_z-\lambda,\lambda\rangle  |^2
        \nonumber \\
  &   & \times
        \sum_{S_z} | \langle JJ_z | J_z-S_z, S_z \rangle  |^2
        \frac{\pi}{M^4} \; \delta \left( 1 - \frac{M^2}{\hat s} \right)
        \overline{\sum_a} \;
        \left| A \left( ij \rightarrow
        c\bar{c} \left[ \mbox{}^3 S_1^{(8)}; S_z, a \right] \right)
        \right| ^2  .
\end{eqnarray}

\begin{figure}
\caption{Gluon-gluon and quark-antiquark flux factors,
$\Phi_{gg}$ and \mbox{$\sum_q (\Phi_{q\bar{q}} + \Phi_{\bar{q}q})$},
plotted as a function of \mbox{$\tau = \hat s / s = x_1 x_2$}
using the GRV-LO parton distributions \protect\cite{GRV}.
(a) $\pi^-$-proton reactions, (b) proton-proton reactions,
(c) antiproton-proton reactions.
Solid line: \mbox{$\Phi_{gg}(\mu_F = M)$}.
Dotted line: \mbox{$\Phi_{gg}(\mu_F = M/2)$}.
Dashed line: \mbox{$\sum_q \left[\Phi_{q\bar{q}} + \Phi_{\bar{q}q}
\right] (\mu_F = M)$}.
Dash-dotted line: \mbox{$\sum_q \left[\Phi_{q\bar{q}} + \Phi_{\bar{q}q}
\right] (\mu_F = M/2)$.} We used \mbox{$M = 3.5$ GeV.}}
\label{flux}
\end{figure}

\mediumtext
\begin{table}
\caption{Leading-twist color-singlet and color-octet contributions
to $J/\psi$ production in $\pi^- N$ collisions at 300 GeV
shown together with experimental data 
\protect\cite{Akerlof,AntoniazziPRL70}.
The cross sections have been integrated over $x_F > 0$ and normalized by
\mbox{$\sigma_{\rm exp}( {\rm all} \; J/\psi) = 178 \pm 21$ nb}.
The dependence of
theoretical contributions on the strong coupling constant $\alpha_s$
and on the relative velocity $v$ of the quark and antiquark in the
bound state is indicated. We used the GRV-LO parton distributions
\protect\cite{GRV}.
}
\begin{tabular}{lccc}
Observed process & & & \\
\hspace{3em} theoretical subprocesses & Scaling
& $\sigma/\sigma_{\rm exp}( {\rm all} \; J/\psi)$ & $\alpha$ \\
\tableline
$\pi^- N \rightarrow \chi_2 \rightarrow J/\psi+\gamma$
& & $0.143 \pm 0.020$ & \\
\hspace{3em} $gg \rightarrow \chi_2$ & $\alpha_s^2 v^5$
& 0.059 & 1.0\tablenotemark[1] \\
\hspace{3em} $q\bar{q} \rightarrow \,^3S_1^{(8)} \rightarrow \chi_2$
& $\alpha_s^2 v^5$ & 0.0041\tablenotemark[2] & 0.29 \\
$\pi^- N \rightarrow \chi_1 \rightarrow J/\psi+\gamma$
& & $0.201 \pm 0.024$ & \\
\hspace{3em}
$q\bar{q} \rightarrow \chi_1 g$ & $\alpha_s^3 v^5$ & 0.0016 & 0.19 \\
\hspace{3em}
$qg \rightarrow \chi_1 q$ & $\alpha_s^3 v^5$ & 0.0029 & -0.22 \\
\hspace{3em}
$gg \rightarrow \chi_1 g$ & $\alpha_s^3 v^5$ & 0.0035 & -0.27 \\
\hspace{3em} $q\bar{q} \rightarrow \,^3S_1^{(8)} \rightarrow \chi_1$
& $\alpha_s^2 v^5$ & 0.0034\tablenotemark[2] & 0.2 \\
$\pi^- N \rightarrow {\rm direct} \; J/\psi$
& & $0.56 \pm 0.03$ & \\
\hspace{3em} $gg \rightarrow J/\psi + g$
& $\alpha_s^3 v^3$ & 0.072 & 0.26 \\
\hspace{3em}
$gg \rightarrow \,^3P_J^{(8)} \rightarrow J/\psi$
& $\alpha_s^3 v^7$ & 0.16\tablenotemark[3] & 0.5 \\
\hspace{3em}
$gg \rightarrow \,^1S_0^{(8)} \rightarrow J/\psi$
& $\alpha_s^3 v^7$ & & 1.0 \\
\hspace{3em}
$q\bar{q} \rightarrow \,^3S_1^{(8)} \rightarrow J/\psi$
& $\alpha_s^3 v^7$ & 0.020\tablenotemark[2] & 1.0 \\
$\pi^- N \rightarrow {\rm all} \; J/\psi$
& & 1 & $0.028 \pm 0.004$ \\
\end{tabular}
\tablenotetext[1]{Reduced to $\alpha \approx 0.85$ if transverse
momentum smearing is taken into account \protect\cite{VHBT}.}
\tablenotetext[2]{Using the color-octet matrix elements of
Ref.\ \protect\cite{ChoII}.}
\tablenotetext[3]{Combined contribution from $^3P_J^{(8)}$ and
$^1S_0^{(8)}$ intermediate states, using the linear combination
of color-octet matrix elements determined in Ref.\ \protect\cite{AFM}.}
\label{table:all}
\end{table}

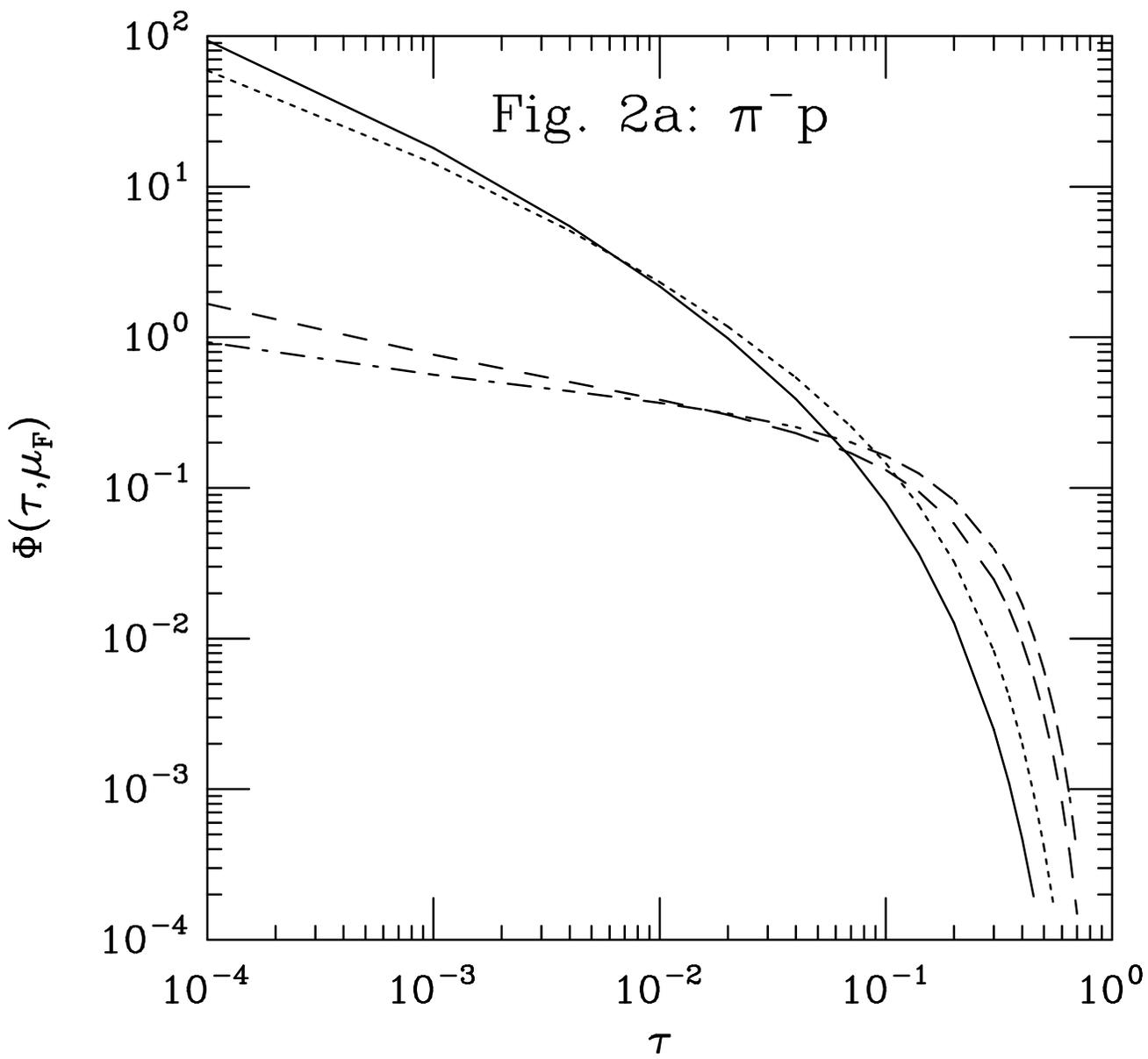
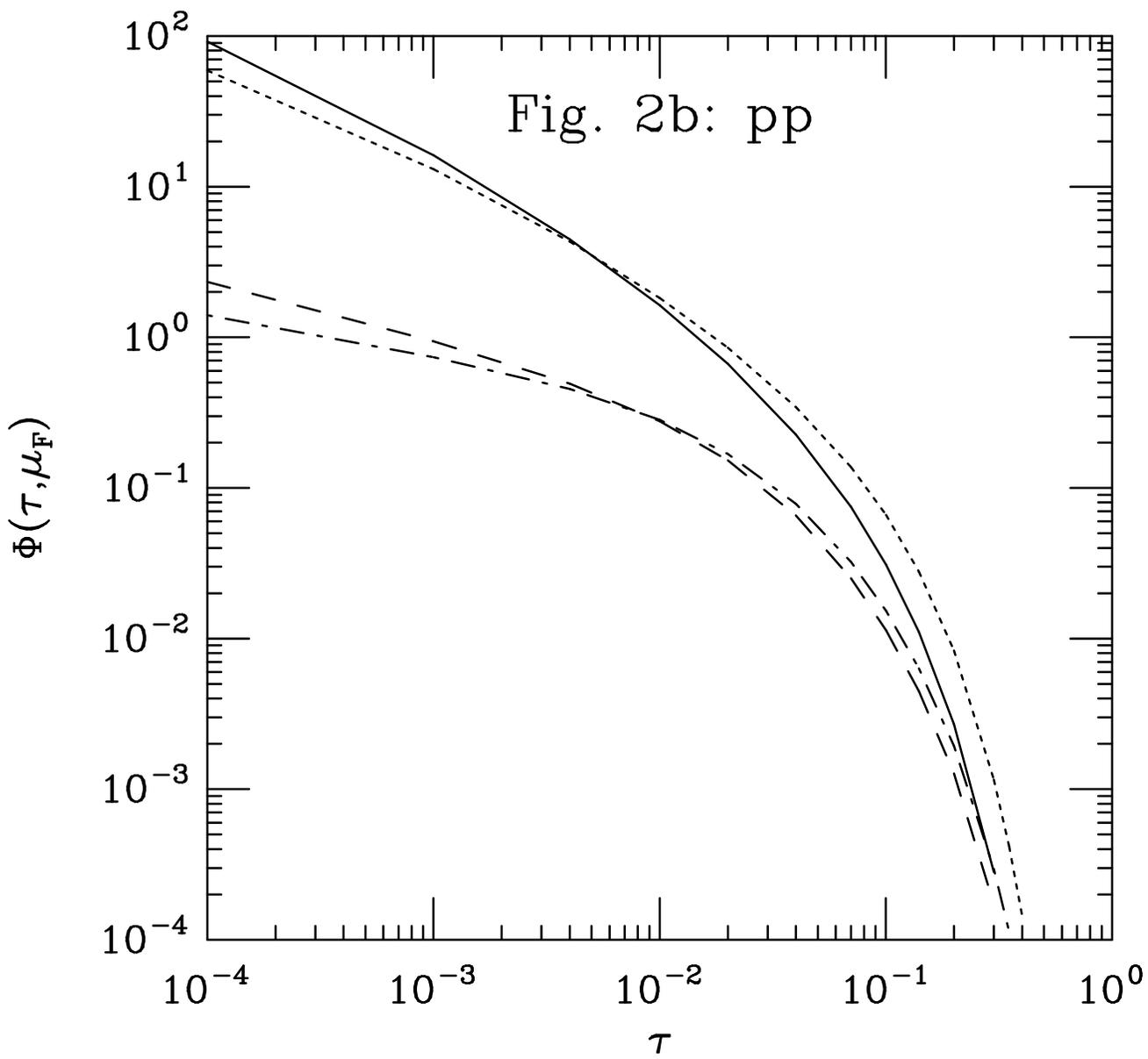
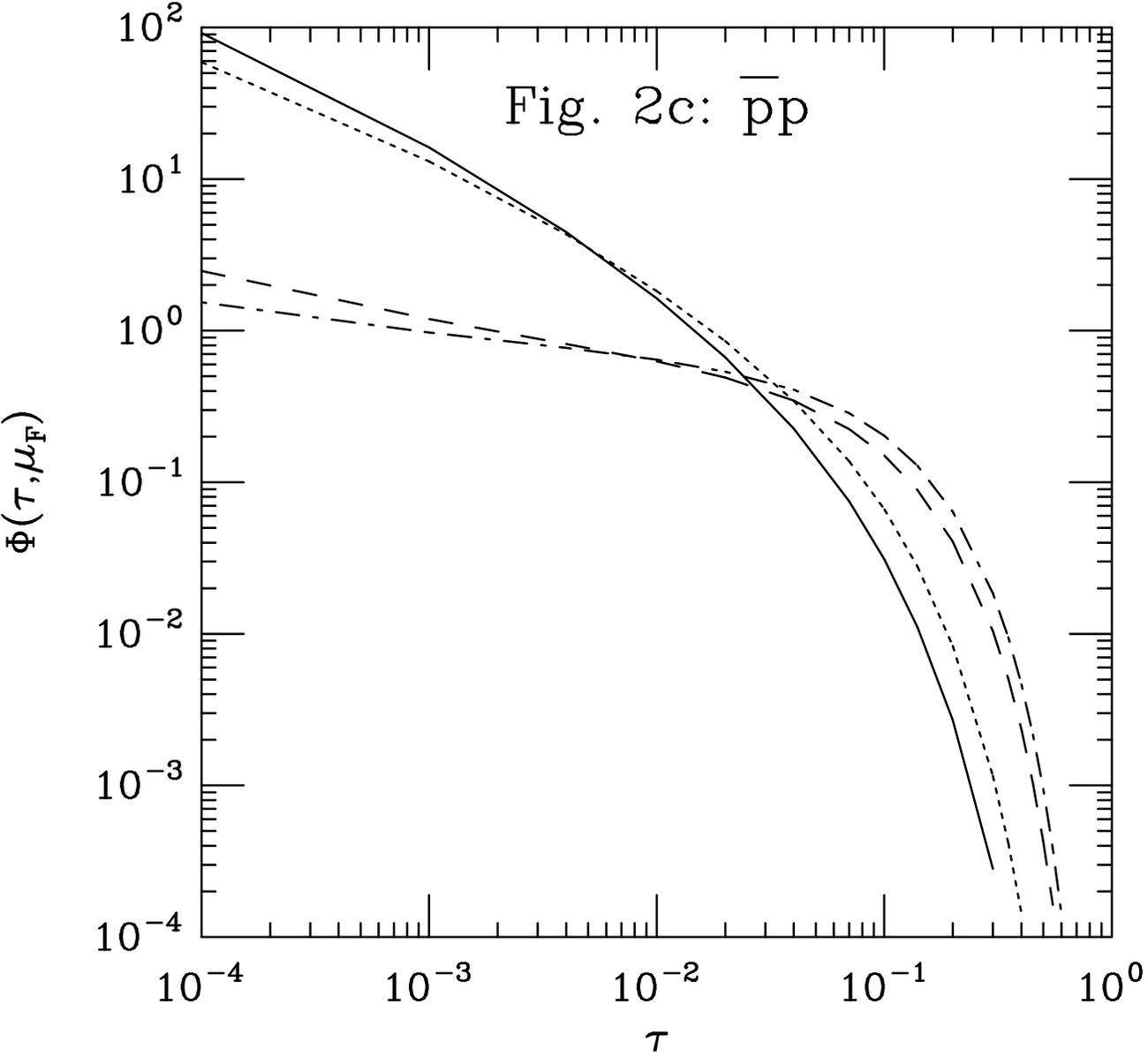
\end{document}